\documentclass[aps,prd,twocolumn,nofootinbib,superscriptaddress]{revtex4-1}

\usepackage{amsmath}
\usepackage{amssymb}
\usepackage{graphicx}

\newcommand{\rfn}[1]{(\ref{#1})}
\newcommand{\gsim}{\lower.7ex\hbox{$\;\stackrel{\textstyle>}{\sim}\;$}}
\newcommand{\lsim}{\lower.7ex\hbox{$\;\stackrel{\textstyle<}{\sim}\;$}}

\begin{document}

%%%%%%%%%%%%%%%%%%%%%%%%%%%%%%%%%%%%%%%%%%%%%%%%%%%%%%%%%%%%%%%%%%%%%%%%%%%%%%%% 
%                               FRONT MATTER
%%%%%%%%%%%%%%%%%%%%%%%%%%%%%%%%%%%%%%%%%%%%%%%%%%%%%%%%%%%%%%%%%%%%%%%%%%%%%%%%

\title{Relaxion Cosmology and the Price of Fine-Tuning}

\author{Stefano Di Chiara}
\affiliation{National Institute of Chemical Physics and Biophysics, R\"avala 10, 10143 Tallinn, Estonia.}
\author{Kristjan Kannike}
\affiliation{National Institute of Chemical Physics and Biophysics, R\"avala 10, 10143 Tallinn, Estonia.}
\author{Luca Marzola}
\affiliation{National Institute of Chemical Physics and Biophysics, R\"avala 10, 10143 Tallinn, Estonia.}
\affiliation{Institute of Physics, University of Tartu, Ravila 14c, 50411 Tartu, Estonia. }
\author{Antonio Racioppi}
\affiliation{National Institute of Chemical Physics and Biophysics, R\"avala 10, 10143 Tallinn, Estonia.}
\author{Martti Raidal}
\affiliation{National Institute of Chemical Physics and Biophysics, R\"avala 10, 10143 Tallinn, Estonia.}
\affiliation{Institute of Physics, University of Tartu, Ravila 14c, 50411 Tartu, Estonia. }
\author{Christian Spethmann}
\affiliation{National Institute of Chemical Physics and Biophysics, R\"avala 10, 10143 Tallinn, Estonia.}

\date{\today}

\begin{abstract}

The relaxion scenario presents an intriguing extension of the standard model in which the particle introduced to solve to the strong CP problem, the axion, also achieves the dynamical relaxation of the Higgs boson mass term. In this work we complete this framework by proposing a scenario of inflationary cosmology that is consistent with all the observational constraints: the relaxion hybrid inflation with an asymmetric waterfall. In our scheme, the vacuum energy of the inflaton drives inflation in a natural way while the relaxion slow-rolls. The constraints on the present inflationary observables are then matched through a subsequent inflationary epoch driven by the inflaton. We quantify the amount of fine-tuning of the proposed inflation scenario, concluding that the inflaton sector severely decreases the naturalness of the theory.
\end{abstract}

\pacs{98.80.Cq, 	% Particle-theory and field-theory models of the early Universe (including cosmic pancakes, cosmic strings, chaotic phenomena, inflationary universe, etc.)
12.60.Fr, 	% Extensions of electroweak Higgs sector
14.80.Va %	Axions and other Nambu-Goldstone bosons (Majorons, familons, etc.)
}

\maketitle

%%%%%%%%%%%%%%%%%%%%%%%%%%%%%%%%%%%%%%%%%%%%%%%%%%%%%%%%%%%%%%%%%%%%%%%%%%%%%%%%
%                               BODY
%%%%%%%%%%%%%%%%%%%%%%%%%%%%%%%%%%%%%%%%%%%%%%%%%%%%%%%%%%%%%%%%%%%%%%%%%%%%%%%%

%-------------------------------------------------------------------------------
\section*{Introduction} % (fold)
\label{sec:Introduction}
%-------------------------------------------------------------------------------

% The Relaxion mechanism
The existence of large hierachies between the Planck scale and other observed scales in Nature poses one of the most perplexing puzzles of contemporary physics. The separation between the Planck scale and the scale of vacuum energy, for instance, spans 31 orders of magnitude and is connected to the famous cosmological constant problem. In order to explain the origin of this separation, in 1984 Abbott \cite{Abbott:1984qf} proposed a relaxation mechanism that recovered the desired vacuum energy value through quantum tunneling, owing to a linear term added to the periodic potential of the QCD axion. 

A similar idea was proposed to explain the origin of the electroweak scale\cite{Dvali:2003br,Dvali:2004tma} and more recently refined in~\cite{Graham:2015cka}. In the ``relaxion'' scenario, the large hierarchy between the Higgs mass scale and the high-energy cut-off of the theory results from the interplay between the dynamics of the QCD axion\footnote{The relaxion mechanism can also be implemented with a generic axion-like particle. We focus, however, on the more attractive possibility in which this new particle has the right properties to solve the strong CP problem of QCD.} and a feedback mechanism arising from the electroweak (EW) symmetry breaking. As the (rel)axion rolls down a linear potential, its coupling to the Higgs boson induces a negative effective mass term. As a result, the Higgs field acquires a non-zero vacuum expectation value (VEV), which breaks the EW symmetry, and a series of increasingly high barriers appears in the relaxion potential. This feedback mechanism stops the rolling of the axion field soon after the onset of the symmetry breaking, leading in a natural way to the generation of the observed EW scale.

% What we do
Since its proposal, the relaxion mechanism has been thoroughly scrutinised and several analyses have improved on the original model \cite{Espinosa:2015eda,Hardy:2015laa,Patil:2015oxa,Antipin:2015jia,Jaeckel:2015txa,Gupta:2015uea,Batell:2015fma,Matsedonskyi:2015xta,Marzola:2015dia,Choi:2015kq,Kaplan:2015fuy,Kobakhidze:2015jya}.
In this work we extend the relaxion framework with a consistent scenario of inflationary cosmology that naturally provides the large number of $e$-folds required by the relaxation mechanism. Our setup complies with observational constraints through a second phase of inflation, which is triggered after the Higgs VEV has stabilized. In this regard, we find that although the relaxion mechanism solves the fine tuning problem at the level of current experimental measurements of the Higgs sector, requiring the compatibility with inflationary cosmology worsens the naturalness of the theory. 

In the following, after briefly reviewing the original model, we detail our inflation scenario and the resulting reheating dynamics. We then discuss the issue of fine-tuning and finally draw our conclusions. 
%-------------------------------------------------------------------------------
\section*{The QCD relaxion} % (fold)
\label{sec:Fine tuning in the relaxion model}
%-------------------------------------------------------------------------------

In the relaxion model~\cite{Graham:2015cka}, the QCD axion $\phi$ interacts with the standard model (SM) Higgs boson $h$ through the effective Lagrangian
\begin{equation}
-\mathcal{L} \supset \frac{1}{2} (M^2 - g \phi) \; h^2 + V(g \phi)  +
\Lambda^4 \cos \frac{\phi}{f},
\label{eq:relax}
\end{equation}
where $M$ is the cut-off scale of the theory and $g$ is a small constant with dimension of mass. We take the relaxion potential, at the leading order, to be
\begin{equation}
V(g \phi) \simeq -M^2 g \phi+ \frac{1}{2}  g^2 \phi^2.
\end{equation}
The energy scale $\Lambda$ that regulates the amplitude of the periodic part of the relaxion potential is linear in the Higgs VEV and the QCD quark condensate:
\begin{equation}
\Lambda^4 \sim v_{h} \; \langle \bar{q} q \rangle .
\end{equation}
Because in the limit $g \to 0$ the periodic discrete shift symmetry of the relaxion field is restored, any value $g \ll M$ can be regarded as technically natural.

The hierarchy between the Higgs boson and the cut-off scale is naturally explained by requiring a large field excursion of $\phi,$
\begin{equation}
  \Delta \phi \gtrsim \frac{M^{2}}{g},
	\label{eq:field_excursion}
\end{equation}
which yields an effective Higgs boson mass that gradually scans all the values from the cut-off scale down to the measured one. In order for the feedback mechanism that stabilises the Higgs boson mass to work, the relaxion  must descend its potential in a slow-roll regime. This is ensured by a first inflation era which, in order to guarantee the required field excursion, must proceed for
\begin{equation}
  N \gtrsim \frac{H^{2}}{g^{2}} \simeq10^{42}
\end{equation}
$e$-folds.

In the original framework~\cite{Graham:2015cka}, the present cosmological constraints are then supposedly satisfied through a second inflation era in a hybrid inflation setup. However, since the barriers in the relaxion potential result from the backreaction of the QCD dynamics, the Hubble constant $H$ during inflation must not exceed the typical QCD scale of about one GeV. This constraint, on top of the required duration of the first inflation era, strongly constrains the mechanism and questions the viability of inflationary cosmology within the QCD relaxion scenario. In the following we propose a hybrid inflation scenario that naturally complies with the requirement posed by the relaxation mechanism and respects, at the same time, the present observational bounds on inflation.

% section Fine tuning in the relaxion model (end)

%-------------------------------------------------------------------------------
\section*{Relaxion cosmology} % (fold)
%\label{sec:A viable inflation scenario}
%-------------------------------------------------------------------------------

\subsection*{A viable inflation scenario}
\label{sec:A viable inflation scenario}

In order to build a working inflation scenario, we extend the potential in \eqref{eq:relax} to a generic polynomial potential which contains all the possible renormalisable interactions with a scalar inflaton $\sigma$:
\begin{equation}
\begin{split}
  V &= \mu_{1 \phi }^{3} \phi + \frac{1}{2} \mu_{\phi}^{2} \phi^{2}
  + \frac{1}{2} \mu_{H}^{2} h^{2}  + \frac{1}{2} \mu_{\phi H} \phi h^{2}
  \\
  &
  + \frac{1}{4} \lambda_{H} h^{4} + \kappa \, h \cos \frac{\phi}{f} + \frac{1}{4} \lambda_{H\sigma} \sigma^{2} h^{2} 
  \\
 &+ \Lambda^{4}_{\sigma} + \mu_{1\sigma}^{3} \sigma + \frac{1}{2} \mu_{\sigma}^{2} \sigma^{2} + \frac{1}{3} \mu_{3\sigma} \sigma^{3}
 + \frac{1}{4} \lambda_{\sigma} \sigma^{4}   \\
  &
  + \frac{1}{2} \mu_{H\sigma} \sigma h^{2}
  + \mu_{\phi \sigma}^{2} \phi \sigma
  + \frac{1}{2} \mu'_{\phi \sigma} \phi \sigma^{2}
  + \frac{1}{2} \mu''_{\phi \sigma} \phi^{2} \sigma
  \\
  &+ \frac{1}{4} \lambda_{\phi\sigma} \phi^{2} \sigma^{2} 
  + \frac{1}{4} \lambda'_{\phi\sigma} \phi \sigma^{3} + \frac{1}{4} \lambda''_{\phi\sigma} \phi^{3} \sigma.
\end{split}
\label{eq:rel:infl}
\end{equation}
Supposing that the relaxion is moving towards positive field values, the relaxation mechanism is implemented for $\mu_{H}^{2} = M^{2}$, $\mu_{\phi H} = -g$, $\mu_{1\phi}^{3} = -M^{2} g$ and $\mu_{\phi}^{2} = g^{2}$. Note that all these terms break the linear shift symmetry of the axion and are assumed to arise from the same source. There will be loop corrections to the terms, but these are of higher order in $g$ and have been neglected.

\begin{figure}[htbp]
\begin{center}
\includegraphics[width=0.48\textwidth]{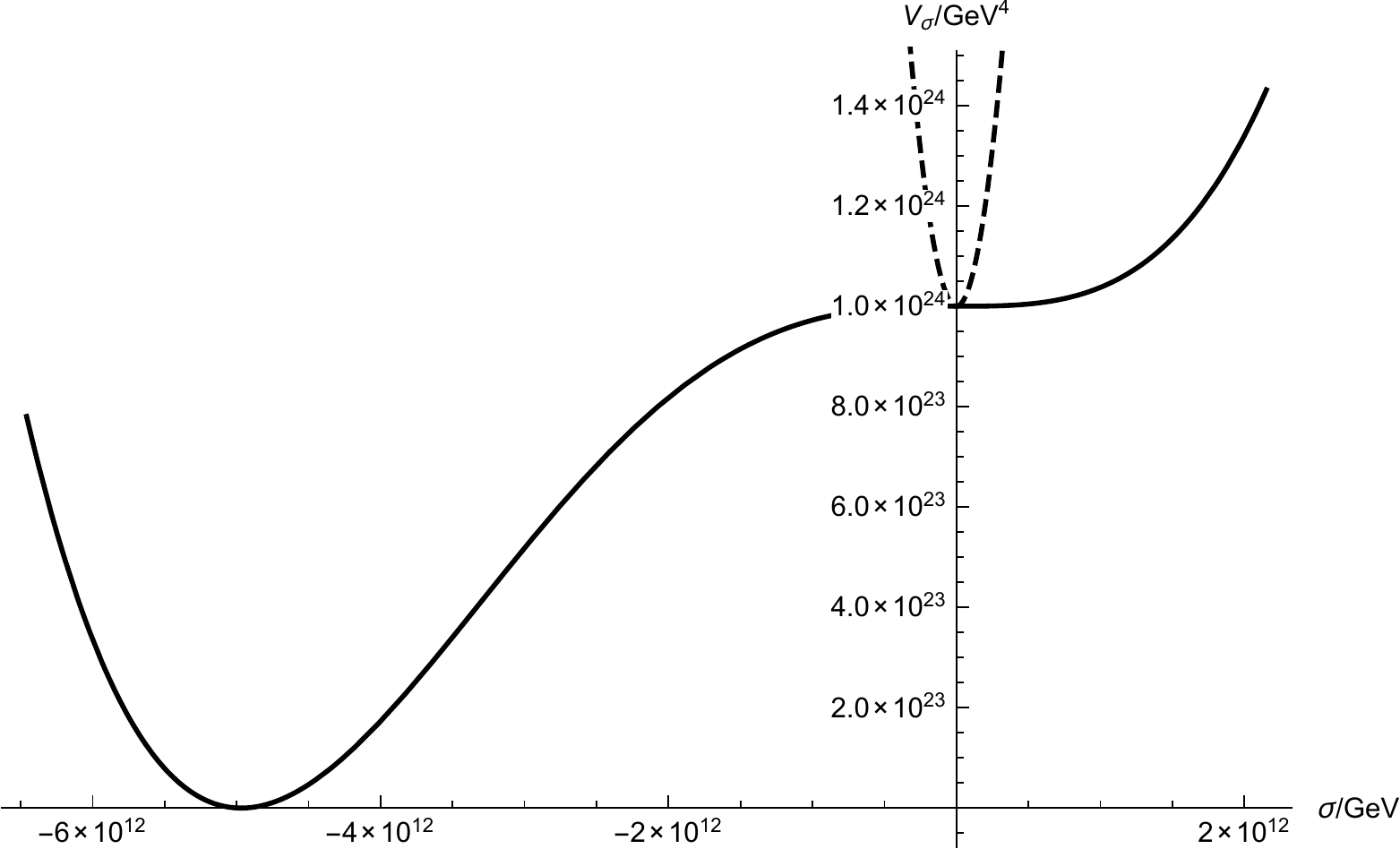}
\caption{The inflaton potential before (dashed line) and after (solid line) the EW symmetry breaking.}
\label{fig:inflaton:V}
\end{center}
\end{figure}

The large number of $e$-folds required by the relaxion mechanism can naturally be produced in a hybrid inflation setup. The dynamics in our mechanism proceeds through the following stages:

\begin{enumerate}

\item \textbf{Relaxation era}\\
In a first phase the $\sigma$ field is confined in minimum of its potential characterised by a non-vanishing potential value (dashed line in Fig.~\ref{fig:inflaton:V}) while the relaxion field is rolling down its potential. The potential energy provided by the $\sigma$ field acts as an effective cosmological constant that drives the inflation dynamics and ensures that the relaxion field is slow-rolling.\footnote{Notice that even in the absence of a third scalar field, the vacuum energy of the Higgs field can sustain hybrid inflation up to the EW symmetry breaking.} 

\item \textbf{EW symmetry breaking}\\
The relaxion field continues to slow-roll until, by effect of the back-reaction due to the QCD barriers triggered by the EW symmetry breaking, it settles in a minimum of its periodic potential. Consequently, the relaxion field acquires a vev $\phi = v_{\phi} \simeq (M^2+\mu^2_{H, \, \mathrm{SM}})/g$ where $\mu^2_{H, \, \mathrm{SM}}$ is the magnitude of the negative Higgs mass term in the SM. This ensures that the resulting Higgs VEV $v_{h}$ matches the measured value.

\item \textbf{Asymmetric waterfall inflation} \\
As in the case of the Higgs boson, the interactions between relaxion and inflaton yields a negative mass term for the inflaton. In fact, once the relaxion acquires a VEV as described before, we have
\begin{equation}
  \mu_{\sigma,\text{eff}}^{2} =
  \mu_{\sigma}^{2} + \mu^\prime_{\phi\sigma} v_{\phi} + \frac{1}{2} \lambda_{\phi\sigma} v_{\phi}^{2}+ \frac{1}{2} \lambda_{H\sigma} v_{h}^{2}
\end{equation}
and the shape of the inflaton potential is consequently modified. The inflaton field, therefore, rolls down from its original position to a new potential minimum $v_{\sigma}$ following an asymmetric waterfall (solid line in Fig.~\ref{fig:inflaton:V}).
\end{enumerate}

We require that the second inflation era respect the following theoretical and experimental constraints: 1) during inflation $H < \Lambda_{\text{QCD}}$; 2) $V(v_{\phi}, v_{h}, v_{\sigma}) \simeq 0$ to account for the observed dark energy density; 3) the number of $e$-folds from $\sigma_N = 0$ to the end of inflation $N \in [50, 60]$~\cite{BenDayan:2009kv}; and observational conditions that 4) amplitude of the spectrum $A_{s} = 2.195 \times 10^{-9}$; and 5) $n_s = 0.968$ at $\sigma_{N}$~\cite{Ade:2015lrj}.

In order to solve the strong CP problem of QCD, the slope of the relaxion potential must decrease to achieve the correct value of the QCD phase $\theta \lesssim 10^{-10}$. In contrast to \cite{Graham:2015cka}, where $g = 10^{-31}$, our inflaton VEV is non-vanishing and, therefore, we give the relaxion potential a larger initial slope $g = 10^{-31}/\theta \simeq 10^{-21}$ and then \emph{subtract} a contribution arising from the interaction with the inflaton and proportional to the VEV of the latter. We set our potential in a way that $\theta = 0$ at the end of the asymmetric waterfall inflation, however the presented results hold also for different values of the QCD phase.

We take the height of the barrier $\kappa \, v_{h} = \Lambda^{4} = 0.1~\text{GeV}^{4}$, the constant $f = 10^{9}~\text{GeV}$ and the cut-off $M = 10^{4}~\text{GeV}$ from \cite{Graham:2015cka}, and set the Higgs quartic coupling  to its SM value $\lambda_{H} = 0.1291$. The values of potential parameters for our reference point are given in Table~\ref{tab:ref:point}. The VEV of the relaxion is $v_{\phi} = 1.0 \times 10^{29}~\text{GeV}$.

\begin{figure}[tbh]
\begin{center}
\includegraphics[width=0.48\textwidth]{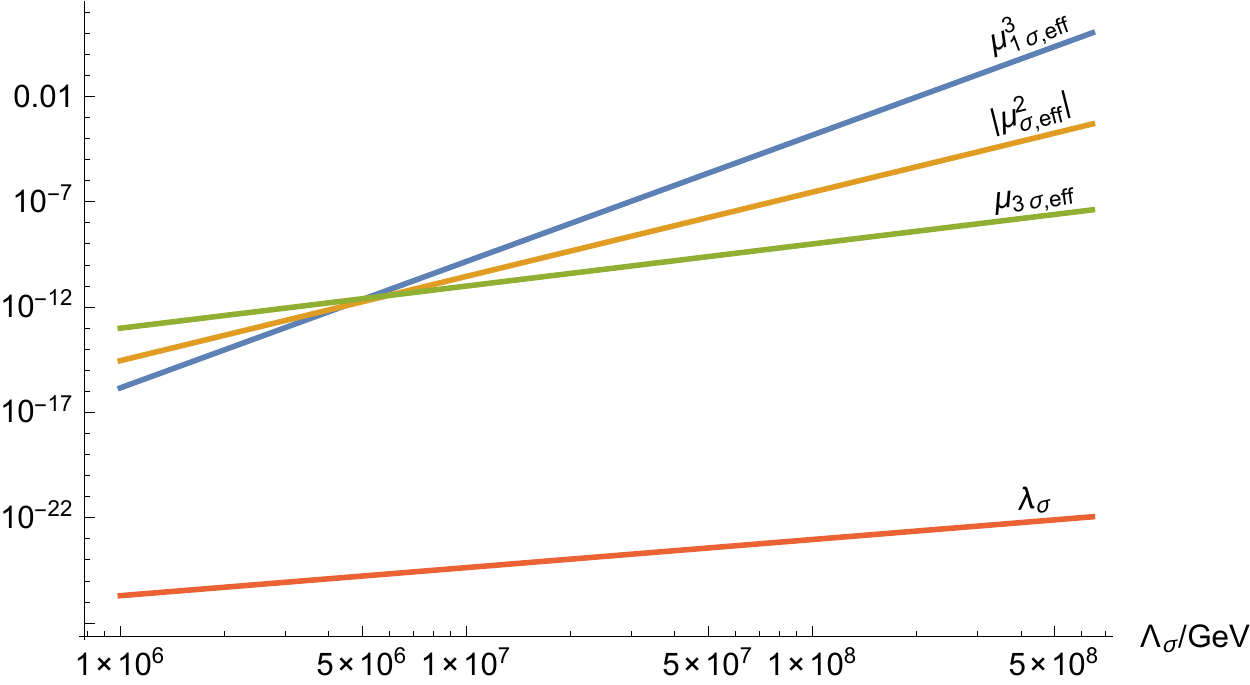}
\caption{The dependence of the inflaton potential parameters on the inflation scale.}
\label{fig:infl:params}
\end{center}
\end{figure}

Setting the inflation scale at $\Lambda_{\sigma} = 10^{6}~\text{GeV}$ and the number of $e$-folds $N = 60$, together with the above constraints, fixes the parameters $\mu_{1\sigma,\text{eff}}^{3}$, $\mu_{\sigma,\text{eff}}^{2}$, $\mu_{3\sigma,\text{eff}}$, and $\lambda_{\sigma}$ in the inflaton potential. Their dependence on the inflation scale is shown in Fig.~\ref{fig:infl:params}.
We choose the value of $\mu_{\sigma}^{2}$ for the inflaton potential to have a minimum at the origin in the symmetric phase and the value of $\lambda_{H\sigma}$ to allow for successful reheating. The resulting inflaton VEV is then  $v_{\sigma} = -4.97 \times 10^{12}$~GeV and its mass is $m_{\sigma} = 0.69~\text{GeV}$. The Hubble constant at $\sigma_{N}$ is $H = 2.4 \times 10^{-7}~\text{GeV}$.

In order to simultaneously satisfy the minimization equations and guarantee the correct values for $\mu_{1\sigma,\text{eff}}^{3}$, $\mu_{\sigma}^{2}$, $\mu_{\sigma,\text{eff}}^{2}$, $\mu_{3\sigma,\text{eff}}$, we solve for  $\mu_{H\sigma}$, $\mu'_{\phi\sigma}$, $\mu_{1\sigma}^{3}$, $\mu_{3\sigma}$, $\lambda_{\phi\sigma}$, taking $\mu_{\phi \sigma}^{2}$, $\mu''_{\phi \sigma}$, $\lambda'_{\phi\sigma}$ and $\lambda''_{\phi\sigma}$ to zero to avoid generation of large linear and cubic terms for the inflaton.\footnote{Notice that the values of inflaton potential parameters given in Table~\ref{tab:ref:point} are approximate, since we cannot present them to the required large number of decimal places.} All the interactions of the inflaton $\sigma$ with the relaxion also break the shift symmetry, but they are required to be small enough already by constraints from inflation itself.

As a result, the predicted tensor-to-scalar ratio is $r \simeq 10^{-42}$, in agreement with the current bound $r < 0.07$~\cite{Array:2015xqh}. We remark that the presented inflationary sector emerges from the most general renormalizable potential allowed in scalar extensions of the relaxion framework. The adopted values of the coefficients have been derived by fitting the current inflationary observables, for instance the spectral index and the spectral amplitude which we set to to their observed values, respectively $n_{s}=0.968$ and $A_{s} = 2.195 \times 10^{-9}$. Therefore, the potential we propose yields {\it the most general} single-field realization of inflationary dynamics that can be achieved through a renormalizable potential, given the constraints imposed by the relaxion framework. Thus, improving on our results would necessarily require an extended particle content or the presence of non-renormalizable interactions in the inflaton potential.

\begin{table}[tb]
\caption{Non-zero potential parameters for our reference point.}
\begin{center}
\begin{tabular}{cc}
Parameter & Value \\
\hline
\hline
$\kappa v_{h}$ & $0.1~\text{GeV}^{4}$
\\
$f$ & $10^{9}~\text{GeV}$
\\
$\mu_{H}^{2}$ & $10^{8}~\text{GeV}^{2}$
\\
$\mu_{\phi H}$ & $-10^{-21}~\text{GeV}$
\\
$\mu_{1\phi}^{3}$ & $-10^{-13}~\text{GeV}^{3}$
\\
$\mu_{\phi}^{2}$ & $10^{-42}~\text{GeV}^{2}$
\\
$\Lambda_{\sigma}$ & $10^{6}$ GeV 
\\
$\mu_{1\sigma,\text{eff}}^{3}$ & $2.84 \times 10^{-16}~\text{GeV}^{3}$
\\
$\mu_{\sigma,\text{eff}}^{2}$ & $-2.77 \times 10^{-15}~\text{GeV}^{2}$
\\
$\mu_{3\sigma,\text{eff}}$ & $9.79 \times 10^{-14}~\text{GeV}$
\\
$\lambda_{\sigma}$ & 1.97 $\times 10^{-26}$
\\
$\mu_{\sigma}^{2}$ & $10~\text{GeV}$
\\
$\lambda_{H\sigma}$ & $10^{-14}$
\\
$\mu_{H\sigma}$ & $0.025~\text{GeV}$
\\
$\mu'_{\phi\sigma}$ & $-2.00 \times 10^{-28}~\text{GeV}$
\\
$\mu_{1\sigma}^{3}$ & $-752.95~\text{GeV}^{3}$
\\
$\mu_{3\sigma}$ & $9.79 \times 10^{-14}~\text{GeV}$
\\
$\lambda_{\phi\sigma}$ & $2.00 \times 10^{-57}$
\\
\hline
\hline
\end{tabular}
\end{center}
\label{tab:ref:point}
\end{table}%

\subsection*{Reheating}

Given the inflaton mass $m_\sigma =  0.69~\text{GeV}$, the reheating can proceed via the decay into electrons, muons, photons and pions through the mixing with the Higgs boson. The reheating temperature is given by
\begin{equation}
  T_{\rm RH} = \left( \frac{90}{g_{*} \pi^{2}} \right)^{\frac{1}{4}} \sqrt{\Gamma_\sigma M_{\rm P}},
\end{equation}
where $g_{*} \sim 100$ is the number of relativistic degrees of freedom and
\begin{equation}
  \Gamma_\sigma \simeq \sin^2 \alpha \ \Gamma_{H}(m_\sigma),
\end{equation}
is the total decay width of the inflaton.
Here $\alpha$  is the mixing angle between the Higgs boson and the inflaton, and  $\Gamma_{H}(m_\sigma)$ is the SM Higgs decay width, computed at a mass $m_\sigma$. We find $\Gamma_{H}(0.69~\text{GeV}) \simeq 2 \times 10^{-8}$~GeV and consequently
\begin{equation}
  T_{\rm RH} \simeq  10^5  \sin \alpha \text{ GeV}.
  \label{TRH}
\end{equation}
The lower bound on the reheating temperature $T_{\text{RH}} > 4.7$~MeV \cite{Dai:2014jja,Munoz:2014eqa,TRHbound} then requires that the portal coupling be necessarily larger than
\begin{equation}
  \sin \alpha \gtrsim 4 \times 10^{-8}.
\end{equation}
For our reference point we obtain $\sin \alpha = 3.91 \times 10^{-4}$.

On the other hand, through the same equation, the current LHC measurements \cite{Cheung:2015dta} result in upper bound on the reheating temperature in our model 
\begin{equation}
	 T_{\rm RH}  \lesssim 7 \times 10^{4} \text{ GeV}
\end{equation}
which, besides successful nucleosynthesis, allows for the generation of the baryon asymmetry of the Universe via neutrino oscillations~\cite{Akhmedov:1998qx}, as well as via resonant leptogenesis \cite{Flanz:1996fb,Covi:1996wh,Pilaftsis:2003gt}. 
%-------------------------------------------------------------------------------
\section*{Fine-tuning} % (fold)
\label{sec:Measuring the fine-tuning}
%-------------------------------------------------------------------------------
We investigate now the impact of the proposed extension on the naturalness of the relaxion mechanism. To this purpose, we first introduce a measure for the fine-tuning and then compute this quantity in both the original model and the proposed extension.

\subsection*{Quantifying the fine-tuning}
Given a set of parameters $\{a_i\}$
%, $i\in\{1,\dots n\}$ 
and a scalar VEV $v(a_i,a_j,\dots)$ related to them by minimization conditions, we quantify the amount of fine-tuning $\Delta_{a_i}$ by~\cite{Barbieri:1987fn,Ellis:1986yg}
\begin{equation}
	\label{ft}
\Delta_{a_i}=\frac{\partial\log v^2(a_i,a_j,\dots)}{\partial\log a_i},
\end{equation} 
corresponding to the ratio of the relative changes in the involved quantities.  

According to the chosen definition, the largest fine tuning of the SM arises from the mass term in the potential
\begin{equation}\label{VSM}
V_{\text{SM}}=\frac{1}{2}\mu_{H}^2 h^2+\frac{1}{4}\lambda_{H} h^4,
\end{equation}
which, including the relevant one-loop corrections, can be written as \cite{Djouadi:2005gi}
\begin{equation}\label{muSM}
\mu_{H}^{2}\left(M\right)=\mu_{H}^{2}\left(m_Z\right)+\frac{3 M^2}{16\pi^2}\left[2\lambda+\frac{g_1^2+3g_2^2}{4}-2y_t^2\right],
\end{equation}
where the tree level contribution is determined by:
\begin{equation}\label{muSMmin}
\partial_h V_{\text{SM}}=0\,\implies \,\mu_{H}^{2}(m_Z)=-\lambda v_h^2 .
\end{equation}
From eqs.~(\ref{ft},\ref{muSM},\ref{muSMmin}) we then find that for the considered cut-off value, $M\simeq 10^4$~GeV, the SM fine tuning amounts to 
\begin{equation}
\label{ftSM}
\Delta_{\mu^2}=328.
\end{equation}
We remark that the fine-tunings of simple extensions of the SM presenting scalar inflaton sectors essentially amount to the fine-tuning of the SM alone. The measured inflationary parameters, in fact, set monomial inflation in its natural energy range of about $10^{13}$ GeV negligible, resulting in a negligible contribution to the fine-tuning of the theory.
We will therefore use the SM value as a reference value in the following discussion.

\subsection*{Local fine-tuning in the relaxion} % (fold)
\label{sub:Fine Tuning of the proposed scenario}

Within the original relaxion framework, the Higgs boson VEV is determined by the interplay between the relaxation mechanism and a feedback effect triggered by the electroweak symmetry breaking. In more detail, after the EW symmetry breaking, the QCD dynamics stop the excursion of the relaxion field\footnote{We neglect the effect of quantum fluctuations which, as pointed out in the original framework, would yield a collection of VEVs having similar values.} in a local minimum of its potential where
\begin{equation}
	g M^2 \approx \frac{\kappa}{f} v_{h},
\end{equation}
with $\kappa$ proportional to the QCD condensate.  According to Eq.~\eqref{ft}, the fine tuning is then
\begin{equation}
	\Delta_g \simeq \Delta_{M^2} \simeq {\cal O}(1),
\end{equation}
confirming the naturalness of the EW scale within the relaxion framework.

Notice, however, that at a ``local'' level the situation is remarkably different.\footnote{By calling it ``local'' we stress that the fine-tuning is calculated in one local minimum, in agreement with the common prescription for fine-tuning.} Suppose that the value of one of the input parameter of the theory, $M$ or $f$ for instance, is modified by such a small amount that the induced change in the final value of the relaxion field is $\delta\phi$,  
\begin{equation}
\label{deltaPhi}
  \delta \phi < 2 \pi f ,
\end{equation}
implying that the relaxion is still stuck in the same local minimum selected by the relaxation mechanism before. Computing now the fine-tuning according to Eq.~\eqref{ft} with respect to the Higgs VEV under the assumption  \rfn{deltaPhi}, we obtain
\begin{equation}
	\Delta_g \simeq \Delta_{M^2} = 12800,
\end{equation}
which exceeds the corresponding value obtained for the SM, Eq.~\eqref{ftSM}, by two orders of magnitude. We interpret the above result as the ``local'' fine-tuning of the relaxion framework: the fine-tuning that the mechanism would have if the experimental value of the Higgs boson VEV had to be matched at a precision comparable with the maximal change induced by $\delta \phi$, quantified by
\begin{equation}
   \delta v_{h} = \frac{g}{v_{h} \lambda} \delta \phi < 10^{-4} \; \mathrm{eV}.
\end{equation}
Our explanation for such a result is that in the relaxion model the cancellation between the large numbers behind the EW scale generation takes place at the tree level, with the tree level parameters being of order of the cut-off scale. In contrast, within the SM, the corresponding cancellation happens at the one-loop level and the same fine-tuning is suppressed by the loop factor. We remark that in the present calculation we adopted the tree-level expression for the parameters in the relaxion model when computing the corresponding fine-tuning. Thus, the value of local fine-tuning that we present could be further enhanced by the contributions of loop corrections. 

Given, however, the precision currently achieved by the EW experiments, the relaxion mechanism is certainly far away from being plagued by its local fine-tuning and the scheme can, indeed, be safely regarded as natural. \newline 

\subsection*{Fine-tuning in the inflaton sector}

The fine-tuning of the parameters introduced by our inflaton sector can be quantified through Eq.~\eqref{ft}. According to our calculation, the largest amount of fine-tuning is due to the squared mass parameter of the inflaton with respect to the relaxion VEV $v_\phi$:
\begin{equation}\label{fti}
\Delta_{\mu_\sigma^2}\simeq 10^{14}\ .
\end{equation}
This large fine-tuning is generated by 
\begin{equation}
\left|\frac{\partial \mu_\sigma^2}{\partial v_\phi}\right|=\mu^\prime_{\phi \sigma }+\lambda _{\phi \sigma } v_{\phi },
\end{equation}
which appears in the denominator of $\Delta_{\mu_\sigma^2}$ and is constrained to be of $O\!\left(g v_h^2 v_\sigma^{-2}\right)$ by the minimization condition of the potential along the $\phi$ direction. An identical amount of fine tuning, evaluated with respect to the inflaton VEV $v_\sigma$, affects the relaxion-inflaton portal coupling $\lambda_{\phi\sigma}$. On top of that, the fine-tuning of the spectral tilt $n_{s}$ with respect to $\Lambda_{\sigma}$ and $\mu_{1\sigma}^{3}$ is $\Delta_{n_{s}} \simeq 10^{14}$, due the inflation scale being much lower than the usual scale of $\mathcal{O}(10^{13})~\text{GeV}$ found within quadratic inflation, for instance.\footnote{If we were to take the Hubble constant close to the highest acceptable level, e.g. $H = 0.1\, \Lambda_{\text{QCD}}$, the fine-tuning of $n_{s}$ would decrease to $\Delta_{n_{s}} \simeq 10^{10}$. Nonetheless, $\Delta_{\mu_\sigma^2}$, which is proportional to $\mu_{\sigma}^{2}$, would increase by about 11 orders of magnitude, since at that scale $\mu_{\sigma}^{2} = 10^{9}$~GeV is necessary to keep the inflaton field at the origin.}

Our conclusion is, therefore, that the required inflation sector is severely fine-tuned. However, the inflation fine-tuning is still significantly lower than the fine-tuning of the cosmological constant within every known inflation scenario. 

Hence, in absence of a valid mechanism to enforce the dissipation of any form of vacuum energy, the significance of the above computation toward the naturalness of the theory could indeed be questioned. If the fine-tuning connected with the inflation sector itself is ignored, our computations show that the global and local-fine tuning of the Higgs and relaxion sector are essentially unmodified by the inflation sector.

% subsection Fine Tuning of the proposed scenario (end)
% section A viable inflation scenario (end)

%-------------------------------------------------------------------------------
\section*{conclusions} % (fold)
\label{sec:conclusion}
%-------------------------------------------------------------------------------

In the present work we extended the relaxion framework with a viable inflationary sector which respects the observational constraints. During the relaxation, the inflaton field is held at a potential minimum with a non-vanishing potential value. This provides the vacuum energy that sustains inflation throughout the relaxation process. Once the relaxation is completed, the inflaton is dragged toward a new potential minimum in an asymmetric waterfall, giving rise to a second inflation era which satisfies the experimental bounds and predicts a vanishing tensor-to-scalar ratio $r\simeq10^{-42}$. In order to look into whether the proposed extension worsens the naturalness of the relaxion mechanism, we investigated the change in fine-tuning due to the additional terms in our potential. We find that whereas the naturalness of the Higgs sector is essentially unaltered, the fine-tuning in the inflaton sector is severe but still lower than the fine-tuning required for the cosmological constant within every inflation scenario. 
% section conclusion (end)

%%%%%%%%%%%%%%%%%%%%%%%%%%%%%%%%%%%%%%%%%%%%%%%%%%%%%%%%%%%%%%%%%%%%%%%%%%%%%%%%
%                               BACK MATTER
%%%%%%%%%%%%%%%%%%%%%%%%%%%%%%%%%%%%%%%%%%%%%%%%%%%%%%%%%%%%%%%%%%%%%%%%%%%%%%%%

\begin{acknowledgments}
The authors thank Matti Heikinheimo for useful discussions. This work was supported by Estonian Research Council grants PUT716, PUT799, PUT1026, IUT23-6 and PUTJD110, by the Estonian Ministry of Education and Research,  
and by the EU through the ERDF CoE program.
\end{acknowledgments}

%%%%%%%%%%%%%%%%%%%%%%%%%%%%%%%%%%%%%%%%%%%%%%%%%%%%%%%%%%%%%%%%%%%%%%%%%%%%%%%%
%                               BIBLIOGRAPHY
%%%%%%%%%%%%%%%%%%%%%%%%%%%%%%%%%%%%%%%%%%%%%%%%%%%%%%%%%%%%%%%%%%%%%%%%%%%%%%%%

\bibliographystyle{apsrev4-1}
\bibliography{relaxion}

\end{document}